\documentclass[aip,jap,reprint,graphicx,floatfix]{revtex4-1} 
\pdfoutput=1

\usepackage{amsmath}
\usepackage{graphicx}

\draft 

\begin{document}



\title{A small mode volume tunable microcavity: development and characterization}



\author{Lukas Greuter}
\author{Sebastian Starosielec}
\author{Daniel Najer}
\author{Arne Ludwig}
\author{Luc Duempelmann}
\author{Dominik Rohner}
\author{Richard J. Warburton}
\affiliation{Department of Physics, University of Basel, Klingelbergstrasse 82, Basel 4056, Switzerland}


\date{4 August 2014}

\begin{abstract}
We report the realization of a spatial and spectrally tunable air-gap Fabry-P\'erot type microcavity of high finesse and cubic-wavelength-scale mode volume. These properties are attractive in the fields of opto-mechanics, quantum sensing and foremost cavity quantum electrodymanics. The major design feature is a miniaturized concave mirror with atomically smooth surface and radius of curvature as low as $10\,\mu\text{m}$ produced by $\text{CO}_2$ laser ablation of fused silica. We demonstrate excellent mode-matching of a focussed laser beam to the microcavity mode and confirm from the frequencies of the resonator modes that the effective optical radius matches the physical radius. With these small radii, we demonstrate sub-wavelength beam waists. We also show that the microcavity is sufficiently rigid for practical applications: in a cryostat at $4\,\text{K}$, the root-mean-square microcavity length fluctuations are below $5\,\text{pm}$.
\end{abstract}

\maketitle 
\thispagestyle{empty}
 
An enhanced interaction between photons and quantum emitters offers a rich field of quantum applications, including single photon transistors and emitter-emitter coupling. Tailoring the vacuum properties of high-Q optical resonators facilitates this enhanced interaction, and ultimately allows a coherent and reversal exchange of energy quanta, the strong coupling regime, challenging to achieve at optical frequencies.~\cite{vahala2003optical} In solid state systems, successful frameworks for cavity quantum electrodynamics (CQED) have been demonstrated with photonic crystal cavities~\cite{HennessyNat2007, YoshieNature2004} and micropillars.~\cite{ReithmaierNature2004} Both approaches allow only very limited in situ tuning.

Upcoming air-gap type resonators such as fiber-based microcavities~\cite{SteinmetzAPL2006,HungerNJP2010,muller2009coupling} offer intrinsic tunability both in the spectral and spatial domains, and continual advancements in thin-film mirror techniques potentially enable ultra-high Q-factors. In this approach, the fiber terminus is fabricated into a concave mirror, where the radius of curvature $\mathcal{R} \approx \mathcal{O}(100\,\mu\text{m})$ defines the resulting cavity mode volume. Such systems however lack precise control over the fiber-to-cavity mode-matching and polarization. In fact, mode-matching is likely to be poor once $\mathcal{R}$ is strongly reduced at which point the beam waist of the cavity mode is substantially smaller than the beam waist of the propagating mode in a typical optical fiber.

We present an open-geometry realization of a miniaturized high-Q Fabry-P\'erot microcavity which allows both spectral and spatial tuning yet overcomes the disadvantages of a fiber-cavity related to mode-matching and polarization. The microcavity is optically accessed by free beam coupling allowing good mode-matching and offers full polarization control in excitation and detection. We present here significant improvements on an earlier approach~\cite{BarbourJAP2011,ZiyunNJP2012}: radii of curvatures down to $10\,\mu\text{m}$ have been achieved and  the microcavity finesse has been significantly enhanced. We demonstrate a sub-wavelength sized beam waists. We measure the frequencies of the microcavity modes in order to determine an effective ``optical" $\mathcal{R}$ and demonstrate that the physical and optical $\mathcal{R}$s match closely. The microcavity has a non-monolithic design and therefore is sensitive to acoustic noise. We quantify the acoustic noise under hostile cryostat conditions. Finally, we estimate relevant CQED parameters for a prototype solid-state emitter, a semiconductor quantum dot, and speculate that large cooperativities can be achieved.

\begin{figure}
\includegraphics[width=85mm]{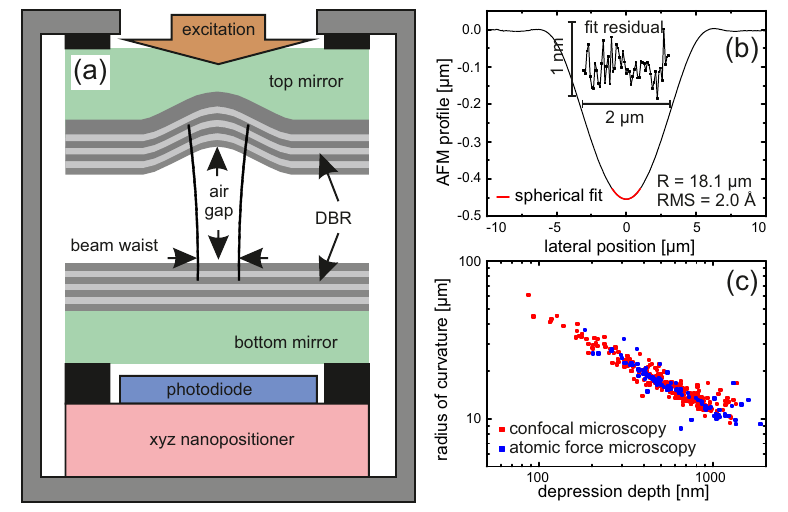} 
\caption{
\label{fig:setup}
(a) The mechanical setup of the concave-planar Fabry-P\'erot air-gap microcavity. A top mirror is produced by $\text{CO}_2$ laser ablation on fused silica and subsequent coating with a high-reflectivity DBR mirror. An opposing bottom mirror with similar reflectivity is mounted on a 3-axis nano-positioner for spectral and spatial tuning. Optical access is through the transparent top mirror substrate; the design of the bottom mirror remains application-dependent.
(b) Profile following ablation of a top mirror template measured with atomic force microscopy. Defect-less near-parabolic craters are achieved with low roughness of $2\,\text{\AA}$. (c) Radius of curvature versus crater depth following ablation. Control of the ablation process parameters gives a wide range of radii of curvatures down to $10\,\mu\text{m}$, with a linear correlation between radius and depth.
}
\end{figure}

A highly reflective planar-concave mirror pair on fused silica, separated by a wavelength-sized air-gap, forms the high-finesse resonator and is shown in figure~\ref{fig:setup}a. Spectral and spatial tunability is achieved by mounting the planar bottom mirror on a 3-axis piezo stack (attocube ANPx/z51, Germany) allowing for sub-nm precise positioning relative to the top mirror. The whole microcavity setup is mounted on an additional 3-axis piezo stack (attocube ANPx/z101) allowing for free positioning with respect to an aspherical lens with $\text{NA}=0.55$ (not shown). This scheme offers control of mode-matching to the microcavity mode~\cite{KogelnikProcIEEE1966} (in contrast to a fiber-cavity) and full polarization control both in excitation and detection. Transmission intensity is measured by a mounted Si photodiode below the bottom substrate. 

For optical characterization we employ different sets of mirror coatings. In a high-reflectivity configuration (HRC), both top and bottom mirror are coated with a $\text{Ta}_2 \text{O}_5 / \text{SiO}_2$ distributed Bragg reflector (DBR) by ion beam sputtering (Evaporated Coatings Inc., USA) with reflectivity reaching $R \approx 99.980\,\%$.  The low-reflectivity configuration (LRC) consists of a $\text{TiO}_2 / \text{SiO}_2$ DBR top mirror coating of proprietary fabrication (OIB GmbH, Germany) with $R > 99.8\,\%$ paired with a low reflective bottom mirror consisting of a polished GaAs substrate coated with a structured $80\,\text{nm}$ Au film ($R \approx 98\,\%$) defined by lift-off electron beam lithography. In both configurations the DBR stopband is centered at $940\,\text{nm}$.

The compact microcavity setup operates both under ambient conditions and also at low temperature. The entire microcavity setup fits into a $50\,\text{mm}$ diameter stainless steel tube containing He exchange gas, the tube is then inserted into a liquid He bath cryostat. The $4\,\text{K}$ conditions favor the mechanical properties of the nano-positioners, reduce photocurrent noise, and demonstrates the low-temperature conformance required for nanostructure systems. The cryostat itself is mounted on an optical table with passive vibrational isolation from the building floor and is situated in a steel chamber damped with acoustic foam.

Prior to coating, the top mirror template is fabricated by laser ablation of fused silica.~\cite{HungerAIPAdv2012,petrak2011feedback} We use an RF-pumped $\text{CO}_2$-laser with wavelength $10.6\,\mu\text{m}$ operating at a repetition rate of $20\,\text{kHz}$. A typical duty cycle of $8\,\%$ results in an average output power of $7.8\,\text{W}$. The light is diffracted by an acousto optical modulator, which allows sufficient control over the incident power and pulse train length in the first order diffracted beam. We reject back-reflection by a linear polarizer and quarter-wave plate. The laser intensity fluctuation in the entire system is less than $1\,\%$. The laser is then focused by an aspherical ZnSe lens ($\text{NA} = 0.45$) onto the silica surface. The silica substrate is mounted on a 3-axis stepper motor stage. In contrast to \citeauthor{HungerNJP2010}.~\cite{HungerNJP2010}, a nitrogen-cooled HgCdTe detector monitors the focal point reflectance as a function of substrate displacement achieving micrometer control over the substrate alignment with respect to the incident laser light. After ablation, the resulting craters are characterized by confocal scanning microscopy and atomic force microscopy. We observe rotationally symmetric micro-craters with a root-mean-square (RMS) surface roughness as low as $2\,\text{\AA}$ (figure~\ref{fig:setup}b). Incident powers of $250\,\text{mW}$ with pulse length between $10 - 100\,\text{ms}$ result in a controlled range of craters with a depth of a few hundred nanometers to $1.5\,\mu\text{m}$ and corresponding radii of curvature down to less than $10\,\mu\text{m}$ (figure~\ref{fig:setup}c). These geometries result from the strong absorption of the $\text{CO}_2$-laser radiation by vibrational modes of the silica, where the melting and evaporation of the material within the first few micrometers of the substrate's surface is roughly proportional to the local intensity. Significantly, the surface tension of the molten silica smooths the ablation craters.

As a central result of this work, ablation craters with radii of curvature down to less than $10\,\mu\text{m}$ are consistently achieved, reducing the previous reported minimum value ($20\,\mu\text{m}$\cite{HungerAIPAdv2012}) by more than a factor of two. We attribute the superior production scheme to an enhanced alignment precision. 

\begin{figure}
\includegraphics[width=85mm]{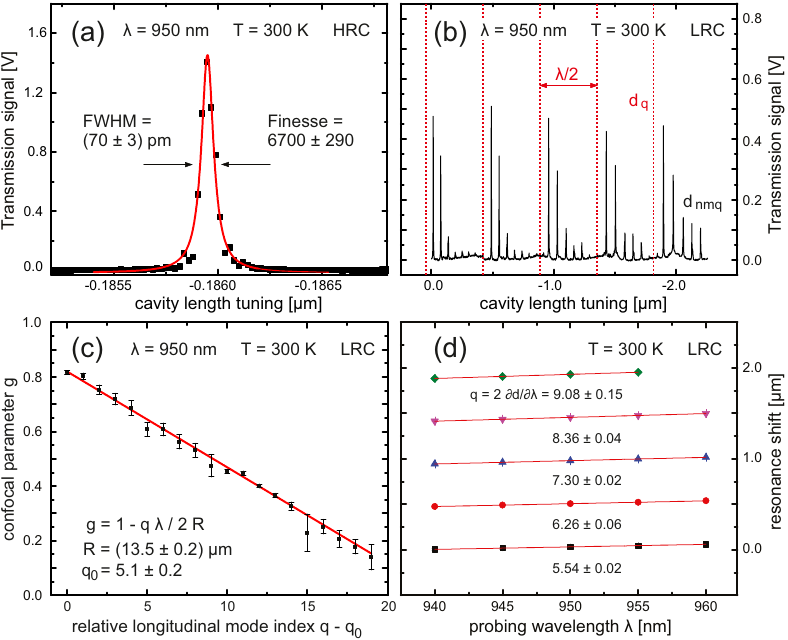} 
\caption{
\label{fig:performance}
Cavity transmission with relative length tuning scans in high-reflectivity configuration (HRC) and low-reflectivity configuration (LRC). (a) An exemplary resonance in HRC. The full-width-at-half-maximum (FWHM) detuning length of $(70 \pm 3)\,\text{pm}$ is measured, corresponding to a finesse of $6{,}700 \pm 290$. (b) Intentionally poor mode-matching in LRC reveals the transverse modes giving a handle on the confocal parameter $g$. (c) The linear relation between $g$ and longitudinal mode index $q$ follows the Gaussian optics model and reveals the top mirror radius of curvature $\mathcal{R} = (13.5 \pm 0.2)\,\mu\text{m}$. The intercept at $g=1$ reveals the longitudinal mode index offset to be $q_0 = 5.1 \pm 0.2$. (d) Independent measurements of the longitudinal mode index for the first five resonances are in very good agreement with $q_0$.
}
\end{figure}

Performance characterization of the microcavity is carried out using transmission detection at a fixed probing wavelength $\lambda = 950\,\text{nm}$. Coarse tuning of the microcavity length is achieved by the friction-based inertial driving mode of the nano-positioner (attocube ANPz51), adjusting the air-gap in $\approx 10\,\text{nm}$ steps over a $2.5\,\text{mm}$ traveling range. For fine tuning over several micrometers, a DC piezo voltage $V_z$ is applied. Figure~\ref{fig:performance}a shows the resonance associated with the fundamental microcavity mode (degenerate with respect to both possible polarizations). There is very good mode-matching: the integrated signal of the fundamental mode is $85\,\%$ of the signal integrated over the fundamental and higher order transverse microcavity modes. To reveal the exact location of the higher order modes, the mode-matching can be made poor intentionally by displacing the entire microcavity setup vertically with respect to the objective lens: Figure~\ref{fig:performance}b shows the transmitted signal as a function of microcavity length tuning, where each peak arises from a particular transverse microcavity mode.

In cavity performance, a representative figure of merit is the finesse usually defined as $F = \Delta \nu_\text{FSR} / \delta \nu$,~\cite{Nagourney} where $\delta \nu$ is the resonance linewidth and $\Delta \nu_\text{FSR}$ is the free spectral range (FSR), i.e.\ the frequency spacing between subsequent longitudinal modes. This definition is ill-defined for DBR-based microcavities as the FSR may become comparable (or even larger) than the DBR stopband. Instead we identify $2\pi / F$ as the fractional energy loss per round-trip at resonance, or $F = c / (2 d \delta \nu)$, where we introduce the microcavity length $d$. On tuning $d$ experimentally, the FWHM resonance width $\delta d$ is immediately accessible. Neglecting the effects of the Gaussian optics (discussed below), we identify $\delta \nu / \nu = \delta d / d$ and measure $F = \lambda / (2 \delta d)$ in the distance domain for fixed vacuum wavelength $\lambda$. Exemplary data in figure~\ref{fig:performance}a show resonance widths of $\delta d = (70 \pm 3)\,\text{pm}$ and thus $F = 6{,}700 \pm 290$ with good lorentzian lineshape (recorded here under ambient conditions). Resonances with FWHM below $30\,\text{pm}$ have been resolved under the same conditions (thus $F \approx 15{,}000$), but their non-lorentzian lineshapes imply a low-frequency broadening as quantified by acoustic noise measurements (also discussed below). In the absence of other cavity losses, reflectivities $R_{1,2} \simeq 99.980\,\%$ as specified for the DBR in HRC, the expected finesse is $F = \pi \sqrt[4]{R_1 R_2} / (1 - \sqrt{R_1 R_2}) \simeq 15{,}700$, in very good agreement with the largest observed values.

We model the mode structure of the microcavity with Gaussian optics. Under the paraxial approximation of optical wavefront normals making only a small angle to the direction of propagation, the propagating electromagnetic field are described by Hermite-Gaussian $\text{TEM}_{nm}$ modes. Imposing two reflective boundaries, the standing wave field is at resonance at mirror separations 
\begin{equation}
\label{eq:resonances}
d_{qnm} = \left[q + \frac{n+m+1}{\pi} \cos^{-1} \sqrt{g} \right] \frac{\lambda}{2} \;,
\end{equation}
where $q$ ($n,m$) is the longitudinal (transverse) resonator mode index and $|g| < 1$ is the geometry-dependent confocal parameter.~\cite{Nagourney} In a planar-concave configuration $g = 1 - d / \mathcal{R}$ depends itself on the mirror separation $d$ and on the concave mirror's radius of curvature $\mathcal{R}$ . The focal point is located at the planar mirror with beam waist $w_0 = \sqrt{\lambda/\pi} \times \sqrt[4]{d\mathcal{R}-d^2}$.

The above analysis holds for idealized mirrors of unity reflection amplitude coefficient. This is not realized in the experiment: the DBR thin-film structures show significant group delay $\tau$ even at the stopband center wavelength.~\cite{BabicIEEE1992} A corresponding effective phase penetration depth $d_\text{DBR} = c / (2\tau)$ of the intracavity field into the DBR affects the resonances by a renormalization of the microcavity length $d_\text{renorm} = d_\text{air-gap} + d_\text{DBR,bottom} + d_\text{DBR,top}$ and radii $\mathcal{R}_\text{renorm}$. The resonance analysis is sensitive to the renormalized microcavity parameters, which in turn are the relevant parameters for CQED applications. We therefore drop the renormalized index in notation and differentiate to the geometrical parameters when needed.

According to eq.~\eqref{eq:resonances}, the splitting of the higher order transverse modes depends on the confocal parameter $g = 1 - d / \mathcal{R}$ and gives a handle on the effective radius of curvature $\mathcal{R}$. An explicit solution of $d$ is cumbersome owing to the implicit nature of eq.~\eqref{eq:resonances} through $g(d)$. We instead exploit the algebraic structure of the splitting and extrapolate $g_q$ at unphysical resonances $m+n+1 \rightarrow 0$ for each longitudinal mode $q$ with $d_q = q \lambda/2$. Figure~\ref{fig:performance}c shows the extracted confocal parameter $g_q$ as a function of integer $q$ with a free offset parameter $q_0$. The Gaussian model relation $g = 1 - q \lambda / (2 \mathcal{R})$ is reproduced for renormalized mirror radius $\mathcal{R} = (13.5 \pm 0.2)\,\mu\text{m}$ and offset $q_0 = 5.1 \pm 0.2$. An independent measurement of the absolute longitudinal mode index $q = 2 \partial d / \partial \lambda$ is performed by changing the probe wavelength (figure~\ref{fig:performance}d) for the first five modes. With the exception of the lowest mode close to the mechanical contact of the mirrors, both the integer spacing as well the offset $q_0 \approx 5.3$ are in very good agreement with the Gaussian optics treatment. In comparison, confocal laser scanning microscopy reveals a geometric radius $\mathcal{R}_\text{geom} = 11.2\,\mu\text{m}$.

\begin{figure}
\includegraphics[width=85mm]{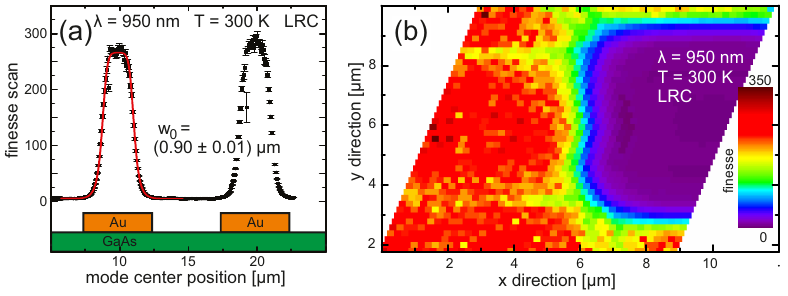}
\caption{
\label{fig:checkerboard}
Cavity mode beam waist measurements using a Au knife-edge structured bottom mirror. (a) A knife edge with $10\,\mu\text{m}$ periodicity for in-situ calibration. A spatial scan of the finesse over knife-edge fits to the model with a beam waist of $w_0 = (0.90 \pm 0.01)\,\mu\text{m}$. (b) Demonstration of two-dimensional microcavity scanning microscopy with a finesse scan over a Au on GaAs panel structure. 
}
\end{figure}

For measurement of the Gaussian beam waist parameter $w_0$ in LRC, the Au bottom mirror is structured with electron beam lift-off lithography into a periodic knife-edge with segment dimensions of $5\,\mu\text{m}$ and periodicity of $10\,\mu\text{m}$. A finesse scan for each position is performed as the edge is moved in the lateral $x$ direction through the microcavity mode. We model the effective reflectivity of the bottom mirror as $R_2^\text{eff}=\alpha R_\text{Au} + (1-\alpha) R_\text{GaAs}$, namely the average reflectivity of the GaAs substrate and Au, where $\alpha(x) = \frac{1}{2}[1 + \text{erf}(\sqrt{2} x / w_0)]$ is the spatial overlap of the Gaussian mode intensity with the Au segments. This simplistic model describes well the observed finesse as a function of bottom mirror position as shown in figure~\ref{fig:checkerboard}a. We determine the beam waist parameter to be $w_0 \approx (0.90 \pm 0.01)\,\mu\text{m}$, equivalently $w_0 / \lambda \approx 0.95 \pm 0.01$:  $w_0$ is smaller than the wavelength. For quantitative comparison of a planar-concave microcavity in Gaussian optics, the beam waist at the focus evaluates to $w_0 = \sqrt{\lambda/\pi} \times \sqrt[4]{d \mathcal{R} - d^2} \approx 1.2\,\mu\text{m}$ for $d \approx 2.5\,\mu\text{m}$ and $\mathcal{R} \approx 13\,\mu\text{m}$, in good agreement with the experimental result.

The above scheme is readily extended to two-dimensional microcavity scanning microscopy, sensitive to the local reflection amplitude coefficient (both in magnitude and phase) with sub-wavelength spatial resolution. A proof of principle is demonstrated in figure~\ref{fig:checkerboard}b for a finesse scan over a Au on GaAs panel structure allowing imperfections in the shape and Au coverage to be detected. 

\begin{figure}
\includegraphics[width=85mm]{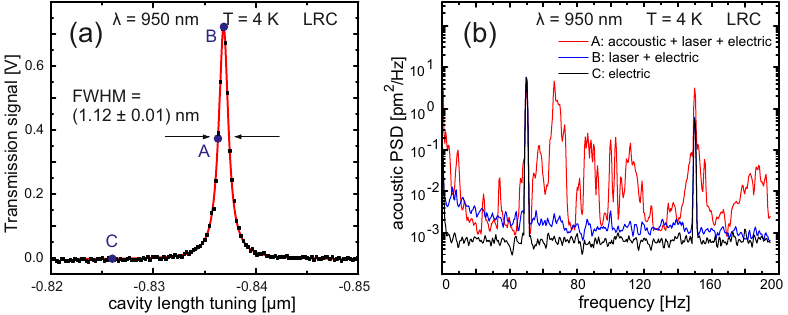}
\caption{
\label{fig:acoustic}
Acoustic noise measurements in the low reflectivity configuration: a fluctuating microcavity length results in a fluctuating resonance position for a constant wavelength. (a) Three working points with sensitivity to different noise sources (see text). (b) Acoustic noise power spectral density at each working point. Red line at A reveals the acoustic noise contribution superimposed on detection noise such as laser intensity noise (blue line, B) and electric noise (black line, C). Noise power is below $1\,\text{pm}/\sqrt{\text{Hz}}$ with the exception of a small band at about $70\,\text{Hz}$. The integrated RMS noise amplitude is $4.3\,\text{pm}$.
}
\end{figure}

We use the strong dependence of the resonance location to the microcavity length as a highly sensitive acoustic noise microphone. Tuning the microcavity to the maximum slope at the resonance edge (working point A in figure~\ref{fig:acoustic}a) gives a linear response in transmission intensity to small air-gap length fluctuations, and hence direct access to the acoustic noise power spectrum affecting the microcavity. Parasitic noise sources such as laser intensity fluctuations and electrical noise are probed at maximum resonance (B) and at significant detuning (C), respectively. The laser intensity is adjusted at each working point to give the same transmitted intensity. The resonance slope in A determines the calibration to acoustic amplitudes and is used in B and C for direct comparison (figure~\ref{fig:acoustic}b). 

Electrical noise (C) sets the detectable equivalent noise floor to $\approx 8\times 10^{-4}\,\text{pm}^2/\text{Hz}$. Mains pick-up noise ($50\,\text{Hz}$ and odd-multiples) is present in the detection channel. From the observed spectral shape in B, laser intensity fluctuations are at most the same order of magnitude as the electrical noise. The acoustic noise spectrum (A) shows rich resonance-like features. However, the acoustic noise amplitudes are still below $10\,\text{pm}^2/\text{Hz}$, and with the exception of the two major contributions at $62-74\,\text{Hz}$ and $150\,\text{Hz}$ well below $1\,\text{pm}^2/\text{Hz}$. Above $200\,\text{Hz}$ no further significant acoustic noise contribution is detected (not shown). The acoustic noise RMS amplitude is $\delta d_\text{acoustic} = 4.3\,\text{pm}$ which must be compared to the FWHM resonance microcavity length $\delta d_\text{resonance} = \lambda / (2 F) = 71\,\text{pm}$ in the HRC setup. If a tolerance of $\delta d_\text{acoustic} / \delta d_\text{resonance} \le 10^{-1}$ is acceptable, a finesse of $F \lesssim 11{,}000$ is unaffected by acoustic noise. Indeed in the experiment, resonances of $\delta d_\text{resonance} = 30\,\text{pm}$ corresponding $F \approx 15{,}000$ consistently feature non-lorentzian lineshapes and in the light of this analysis, this arises most likely from the acoustic noise.

In conclusion, we have demonstrated the experimental realization of a fully tunable open-gap Fabry-P\'erot microcavity. The microcavity modes have small volume and are well described with Gaussian optics. We achieve excellent mode-matching to the microcavity mode from a propagating Gaussian beam. The major facilitators are, first, the successful fabrication of atomically smooth, small radius of curvature top mirrors by $\text{CO}_2$ laser ablation and second, a rigid piezo-driven nano-positioning system. 

We speculate further on the cooperativity that can be reached with our setup when operated with an InGaAs self-assembled quantum dot in a GaAs host matrix grown on a GaAs/AlGaAs DBR. For this configuration, due to the high phase penetration depth into the DBR, the renormalized microcavity length is estimated to be $8.5\,\mu\text{m}$. A Gaussian optics estimate yields an average beam waist of $2.0\,\mu\text{m}$. We use the transfer matrix method to calculate a vacuum electric field amplitude of $E_\text{vac} = 2.0 \times 10^4\,\text{V}/\text{m}$ at the location of the quantum dot. A typical free space radiative lifetime of $0.8\,\text{ns}$ corresponds to an optical dipole moment of $\mu_{12} =  1.2\,\text{nm} \times e$ resulting in an emitter-cavity interaction energy of $\hbar g = \mu_{12}E_\text{vac} = 24\,\mu\text{eV}$. The demonstrated finesse of $6{,}700$ translates into a photon decay of $\hbar\kappa = 22\,\mu\text{eV}$. Quantum dot linewidths as low as $\hbar\gamma= 2\,\mu\text{eV}$ are routinely achieved, resulting in an upper limit cooperativity as high as $C = 2 g^2 / (\kappa \gamma) \approx 26$. Further reduction in $\kappa \ll g$ by supermirror DBR coatings ($R > 99.995\%$) may even enhance the cooperativity towards $100$.

\begin{acknowledgments}
The authors thank M.\ Montinaro who performed the electron-beam lithography of the checkerboard calibration sample. The authors gratefully acknowledge financial support by SNF and NCCR QSIT. 
\end{acknowledgments}

\bibliography{references}

\end{document}